\documentclass[twocolumn]{aastex63}

\newcommand{\target}{COOL\,J0335$-$1927}
\newcommand{\z}{\textit{z}}
\newcommand{\Lenstool}{{\tt{Lenstool}}}
\newcommand{\qsoOne}{SDSS\,J1004$+$4112}
\newcommand{\qsoTwo}{SDSS\,J1029$+$2623}
\newcommand{\qsoThree}{SDSS\,J2222$+$2745}
\newcommand{\qsoFour}{SDSS\,J0909$+$4449}
\newcommand{\qsoFive}{SDSS\,J1326$+$4806}
\newcommand{\qsoSix}{COOL\,J0542$-$2125}

\newcommand{\kms}{km s$^{-1}$}
\newcommand{\qsoz}{3.27}
\newcommand{\clusterz}{0.4178}

\newcommand{\grz}{\textit{grz}}
\newcommand{\predAB}{$499^{+141}_{-146}$}
\newcommand{\predAC}{$-127^{+83}_{-17}$}

\usepackage{comment}
\usepackage{breqn}
\usepackage{amsmath}
\usepackage{longtable}

\shorttitle{}
\shortauthors{Napier et al.}
\graphicspath{{./}{figures/}}
\begin{document}

\title{COOL-LAMPS. Discovery of COOL\,J0335$-$1927, a Gravitationally Lensed Quasar at $z$=3.27 with an Image Separation of 23$\farcs$3}

\correspondingauthor{Kate Napier}
\email{kanapier@umich.edu}

\author[0000-0003-4470-1696]{Kate Napier}
\affiliation{Department of Astronomy, University of
Michigan, 1085 S University Ave, Ann Arbor, MI 48109, USA}

\author[0000-0003-1370-5010]{Michael D. Gladders}
\affiliation{Department of Astronomy and Astrophysics, University of Chicago, 5640 South Ellis Avenue, Chicago, IL 60637, USA}
\affiliation{Kavli Institute for Cosmological Physics, University of Chicago, 5640 South Ellis Avenue, Chicago, IL 60637, USA}

\author[0000-0002-7559-0864]{Keren Sharon}
\affiliation{Department of Astronomy, University of
Michigan, 1085 S University Ave, Ann Arbor, MI 48109, USA}

\author[0000-0003-2200-5606]{H{\aa}kon Dahle}
\affiliation{Institute of Theoretical Astrophysics, University of Oslo, P.O. Box 1029, Blindern, NO-0315 Oslo, Norway}

\author[0000-0001-9978-2601]{Aidan P. Cloonan}
\affiliation{Department of Astronomy and Astrophysics, University of
Chicago, 5640 South Ellis Avenue, Chicago, IL 60637, USA}

\author[0000-0003-3266-2001]{Guillaume Mahler}
\affiliation{Centre for Extragalactic Astronomy, Durham University, South Road, Durham DH1 3LE, UK}
\affiliation{Institute for Computational Cosmology, Durham University, South Road, Durham DH1 3LE, UK}

\author[0000-0003-0896-8502]{Isaiah Escapa}
\affiliation{Department of Astronomy and Astrophysics, University of Chicago, 5640 South Ellis Avenue, Chicago, IL 60637, USA}

\author[0009-0007-1440-1832]{Josh Garza}
\affiliation{Department of Astronomy and Astrophysics, University of Chicago, 5640 South Ellis Avenue, Chicago, IL 60637, USA}

\author[0009-0006-7664-877X]{Andrew Kisare}
\affiliation{Department of Astronomy and Astrophysics, University of Chicago, 5640 South Ellis Avenue, Chicago, IL 60637, USA}

\author[0000-0002-5825-7795]{Natalie Malagon}
\affiliation{Department of Astronomy and Astrophysics, University of Chicago, 5640 South Ellis Avenue, Chicago, IL 60637, USA}

\author[0000-0002-5573-9131]{Simon Mork}
\affiliation{Department of Astronomy and Astrophysics, University of Chicago, 5640 South Ellis Avenue, Chicago, IL 60637, USA}

\author[0009-0001-5944-5624]{Kunwanhui Niu}
\affiliation{Department of Astronomy and Astrophysics, University of Chicago, 5640 South Ellis Avenue, Chicago, IL 60637, USA}

\author[0000-0001-7905-2134]{Riley Rosener}
\affiliation{Department of Astronomy and Astrophysics, University of Chicago, 5640 South Ellis Avenue, Chicago, IL 60637, USA}

\author[0009-0000-9780-4328]{Jamar Sullivan Jr.}
\affiliation{Department of Astronomy and Astrophysics, University of Chicago, 5640 South Ellis Avenue, Chicago, IL 60637, USA}

\author[0000-0002-6832-4680]{Marie Tagliavia}
\affiliation{Department of Astronomy and Astrophysics, University of Chicago, 5640 South Ellis Avenue, Chicago, IL 60637, USA}

\author[0009-0008-0518-8045]{Marcos Tamargo-Arizmendi}
\affiliation{Department of Astronomy and Astrophysics, University of Chicago, 5640 South Ellis Avenue, Chicago, IL 60637, USA}

\author[0000-0002-5279-0230]{Raul Teixeira}
\affiliation{Department of Astronomy and Astrophysics, University of Chicago, 5640 South Ellis Avenue, Chicago, IL 60637, USA}

\author[0009-0008-6557-2065]{Kabelo Tsiane}
\affiliation{Department of Astronomy and Astrophysics, University of Chicago, 5640 South Ellis Avenue, Chicago, IL 60637, USA}

\author[0000-0003-0295-875X]{Grace Wagner}
\affiliation{Department of Astronomy and Astrophysics, University of Chicago, 5640 South Ellis Avenue, Chicago, IL 60637, USA}

\author[0000-0001-6454-1699]{Yunchong Zhang}
\affiliation{Department of Astronomy and Astrophysics, University of Chicago, 5640 South Ellis Avenue, Chicago, IL 60637, USA}

\author[0009-0006-4143-1159]{Megan Zhao}
\affiliation{Department of Astronomy and Astrophysics, University of Chicago, 5640 South Ellis Avenue, Chicago, IL 60637, USA}



\begin{abstract}
%250 words 

We report the discovery of  \target, a quasar at \textit{z} = \qsoz\ lensed into three images with a maximum separation of $23\farcs 3$ by a galaxy cluster at \textit{z} = \clusterz.  To date this is the highest redshift wide-separation lensed quasar known.  In addition, \target\, shows several strong intervening absorbers visible in the spectra of all three quasar images with varying equivalent width. The quasar also shows mini-broad line absorption. We construct a parametric strong gravitational lens model using ground-based imaging, constrained by the redshift and positions of the quasar images as well as the positions of three other multiply-imaged background galaxies.  Using our best-fit lens model, we calculate the predicted time delays between the three quasar images to be $\Delta$t$_{AB}=$ \predAB\ (stat) and $\Delta$t$_{AC}=$ \predAC\ (stat) days. Folding in systematic uncertainties, the model-predicted time delays are within the ranges $240 < \Delta$t$_{AB} < 700$ and $-300 < \Delta$ t$_{AC} <-30$.  We also present \textit{g}-band photometry from archival DECaLS and Pan-STARRS imaging, and new multi-epoch observations obtained between September 18, 2022 UT and February 22, 2023 UT, which demonstrate significant variability in the quasar and which will eventually enable a measurement of the time delay between the three quasar images.  The currently available light curves are consistent with the model-predicted time delays.  This is the fifth paper from the COOL-LAMPS collaboration.

% 250 word limit for the abstract
\end{abstract}



\section{Introduction} \label{sec:intro}

To date, there are more than 300 quasars known to be gravitionally lensed into multiple images \citep{lemon2019, lemon2023}.  Most of these quasars are lensed by individual galaxies, resulting in separations of $1-2$ arcseconds between the lensed quasar images.  There are only six published wide-separation lensed quasars (WSLQs), strongly-lensed quasars in which the foreground lens is a group or cluster of galaxies, producing image separations $>10$ arcseconds: \qsoOne\ \citep{Inada2003}; \qsoTwo\ \citep{Inada2006}; \qsoThree\ \citep{Dahle2013}; \qsoFour\ \citep{Shu2018}; \qsoFive\ \citep{Shu2019}; and \qsoSix\ \citep{Martinez2023}.   

Despite their rarity, WSLQs offer rich astrophysical insights due to the image separations and time delays that can be a magnitude larger than those produced by galaxy lenses. For example, the presence of a leading image $\sim$700 days ahead of the brightest images in SDSS J2222+2745 allowed \citet{Williams2021a,Williams2021b} to plan and execute a reverberation mapping campaign on this \textit{z} = 2.801 WSLQ, providing the best constrained black hole mass currently available in the distant Universe.  WSLQs can be used to study the quasar host galaxy \citep{Ross2009, Cloonan2023}, and interactions between the quasar and nearby gas, at smaller scale than otherwise achievable \citep{Bayliss2017}.  Since each lensed quasar image probes a slightly different line of sight, WSLQs can be used to construct the 3D spatial distribution of outflows \citep{Misawa2013}.  WSLQs uniquely probe the mass distribution of the lens on a variety of scales \citep{Oguri2010, Oguri2013, Liesenborgs2009, Sharon2017, Fores-Toribo2022}.  Recently, \cite{Napier2023} presented a time-delay measurement of the Hubble parameter, H$_{0}$, based on the three best-studied WSLQs.       

In this Letter, we present the discovery, spectroscopic confirmation, preliminary lens model, and photometric monitoring of \target, a quasar at \z\ = \qsoz\ strongly lensed by a galaxy cluster at \z\ = \clusterz\ into three images with maximum image separation of 23\farcs 3.  Of all the known cluster-lensed quasars, this is the highest quasar redshift to date.  

In Section~\ref{sec:obs}, we present the ground-based spectroscopy which confirms this target as a cluster-lensed quasar.  We present an initial strong lensing model in Section~\ref{sec:lens_model}.  In Section~\ref{sec:time_delay}, we present a prediction of the time delays between the quasar images.  In Section~\ref{sec:photometry}, we present preliminary quasar light curves constructed from archival data and our new photometric monitoring campaign.  We conclude in Section~\ref{sec:conclusions}.   

We assume a flat cosmology with $\Omega_{\Lambda} = 0.7$, $\Omega_{m}=0.3$, and $H_0 = 70$ \kms\ Mpc$^{-1}$. In this cosmology, $1''=$ 5.515 kpc at the cluster redshift, $z=$ \clusterz, and $1''=$ 7.495 kpc at the source redshift, $z=$ \qsoz. Magnitudes are reported in the AB system unless otherwise stated.

\section{Discovery}\label{sec:discovery}
\target\ was discovered serendipitously by the COOL-LAMPS collaboration, as part of a systematic visual search for strong lenses \citep[e.g.,][]{Khullar2021,Sukay2022,Zhang2022} in public imaging data from the Dark Energy Camera Legacy Survey \citep[DECaLS:][]{Dey2019}.  It was not flagged as a candidate WSLQ by the process described in \citet{Martinez2023} because two of the three lensed quasar images (images B and C) in \target\ are classified in the DECaLS DR9 catalog as extended sources (S\'ersic profiles). Much like the WSLQ SDSS\,J2222$+$2745 \citep{Dahle2013}, this system was found first due to the presence of arc-like features in survey images, which focused sufficient attention on the images to suggest the presence of three quasar images, and trigger follow-up spectroscopy in September 2022. 

\section{Data}\label{sec:obs}
\subsection{Description of the Observations}
We observed \target\ on September 17, 2022 with the LDSS3-C spectrograph on the Magellan~2 6.5-m Clay Telescope, with a total of three different long-slit positions, each time using the $1\farcs0\times4'$ center long-slit dispersed by the VPH-All grism.  The first and second slit positions, targeting the three candidate lensed quasar images, were observed for 2$\times$900s. The third slit position, targeting multiple apparent cluster lens galaxies, was observed for 900s. Spectra were sky-subtracted, and wavelength and flux calibrated using standard techniques in IRAF
%\footnote{Standard IRAF routines in the iraf.noao.imred.ccdred and iraf.noao.onedspec packages} 
\citep{Tody1986,Tody1993}.  The seeing at the time of spectroscopic observations was $\sim$1\arcsec. 

On the same night we also acquired \grz\ imaging, totaling 600s in the \textit{z} filter, and 540s in the \textit{r} and \textit{g} filters. We obtained additional \grz\ imaging of \target\ on January 31, 2023, once again with the LDSS3-C instrument, with total integration times of 1200s in the \textit{z} filter, and 540s in the \textit{r} and \textit{g} filters. The final stacked \textit{grz} images have seeing of 0$\farcs$85, 0$\farcs$65, and 0$\farcs$55, respectively.    A color image from these data is shown in Figure~\ref{fig:imaging}.  Notably the two quasar images classified as extended in the DECaLS catalogs are not obviously extended in these sharper and deeper data.

\subsection{Results from Spectroscopy}
\begin{figure*}
    \centering
    \includegraphics[width=500px]{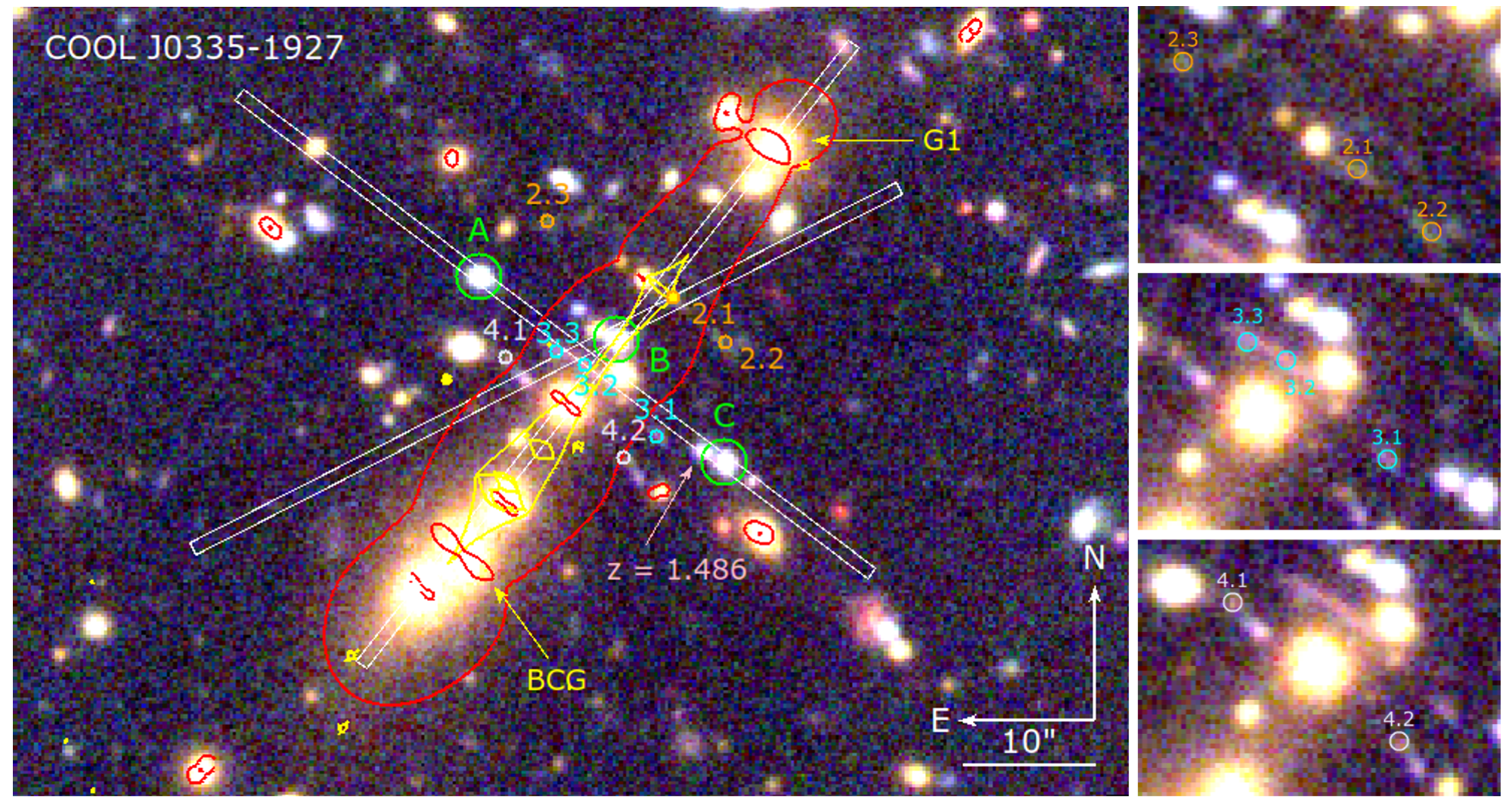}
    \caption{\grz\ color composite imaging of \target\ from Magellan Telescopes LDSS3-C.  The three quasar images are marked in green.  The lensing critical curves, the points of maximum magnification for a source at a specified redshift, are computed at the quasar's redshift \textit{z} = \qsoz\ and shown in red.  Several other multiply-imaged background sources are seen in the vicinity of the quasar images and are shown in the zoomed-in panels on the right.  The white boxes mark the three LDSS3-C long slit observations.}
    \label{fig:imaging}
\end{figure*}

\begin{figure*}
    \centering
    \includegraphics[width=500px]{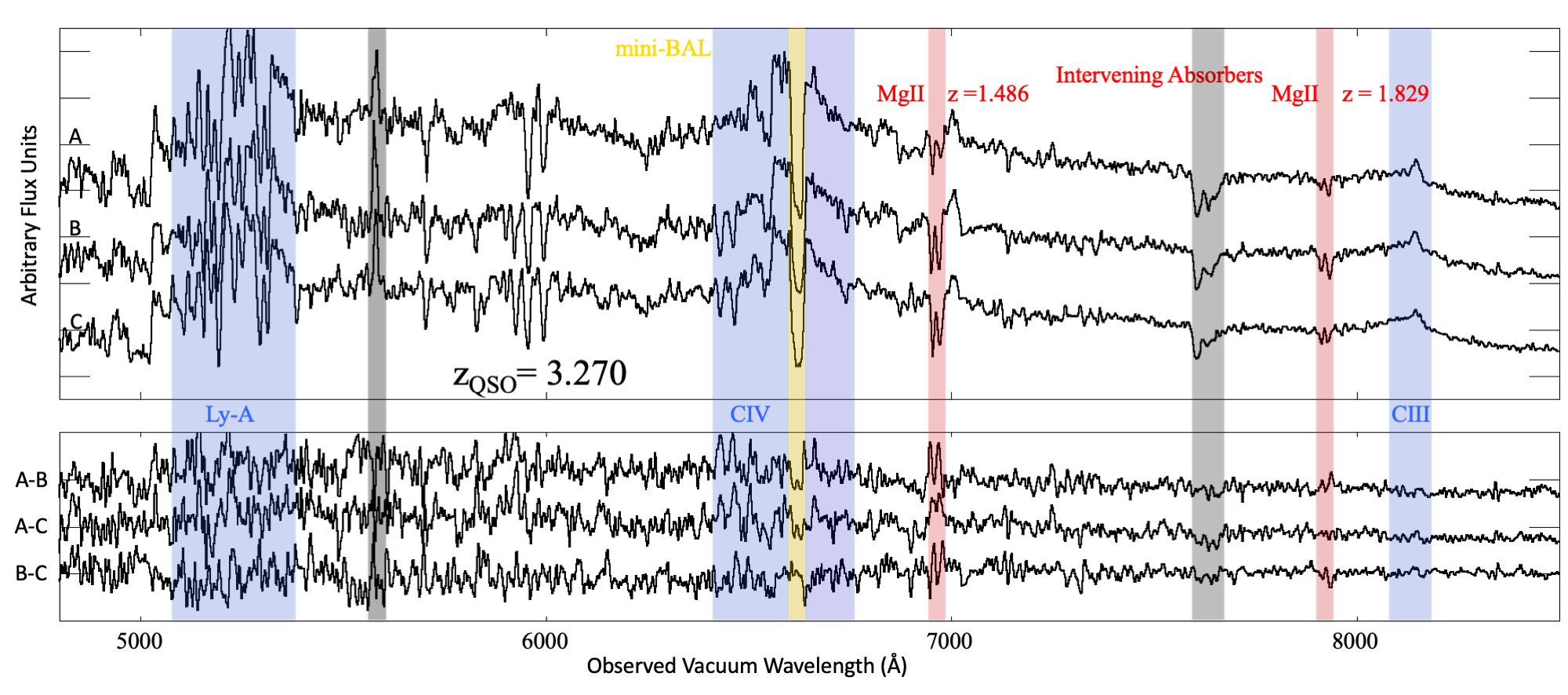}
    \caption{LDSS3-C spectra of the three quasar images of \target.  The spectra are vertically offset for clarity, with the top, middle, and bottom spectra corresponding to images A, B, and C, respectively.  The gray shaded regions indicate major telluric and night sky features.  The C III blue region was used to derive the systemic redshift.  The system is classified as a ``mini"-BAL based on the broad and nearly opaque absorption seen redshifted relative to systemic at CIV.  The red shaded regions indicate a Mg II absorption feature used to identify intervening absorbers at $z=$ 1.486 (see also Figure \ref{fig:imaging}) and at $z=$ 1.829.  The bottom panel shows the difference spectra of the three pairs of quasar images.  The difference spectra are consistent with the noise.  Future work will investigate any differences in the quasar spectra resulting from variability or line-of-sight differences due to intrinsic and intervening absorption.}
    \label{fig:spectra}
\end{figure*}

The LDSS3-C spectra of the three quasar images confirm that they are indeed three images of the same background source, as shown in Figure~\ref{fig:spectra}. 
Based on the He I 1640 and C III lines, we derived a systemic quasar redshift of \z\ = \qsoz, to date the highest redshift of a quasar lensed by a galaxy cluster.
The LDSS3-C R$\sim$550 spectra (at 6000\AA) with a continuum SNR of $\sim$20 per resolution element show that all three quasar images have a significant number of absorption lines. Of particular note is a wide ($\sigma\sim700$ \kms) and almost opaque absorption feature 290 \kms\ redward of the systemic velocity
that classifies \target\ as a ``mini"-BAL (broad absorption line) quasar \citep[e.g.,][]{Barlow1997}.  There are also two foreground absorbers visible in the extant spectroscopy, at z=1.486 and z=1.829, the former of which corresponds to a galaxy in close projection to image C at the same redshift with a detected [OII] doublet in emission (see Figure \ref{fig:imaging}). The three widely-separated lensed quasar images each probe both the absorbing material around the quasar itself and the intervening absorbers along different lines of sight, separated by impact parameters of tens of kpc for the foreground absorbers, and on sub-parsec scales for the intrinsic absorption.  Even at this modest resolution there are apparent differences in equivalent width between the foreground absorbers across the three quasar images - see for example the indicated foreground absorber at z=1.486 in Figure~\ref{fig:spectra}; a more detailed analysis awaits higher resolution spectroscopy. 

\section{Lens Model}\label{sec:lens_model}
We computed an initial strong lensing model for the foreground galaxy cluster lens of \target\ from ground-based imaging and spectroscopy, using as constraints the positions of multiple images of lensed background sources (Table~\ref{table:table}), and where available, their spectroscopic redshifts (currently, only the quasar).     

The strong lensing model presented here was constructed using \Lenstool\ \citep{Jullo2007}.  This parametric modeling algorithm models the lens plane as a linear combination of mass halos.  The mass halos represent the dark matter mass distribution from the cluster-scale to the individual galaxy-scale.  We modeled each mass halo as a pseudo-isothermal ellipsoidal mass distribution \citep[PIEMD or dPIE;][]{Eliasdottir2007}, which has seven parameters: position (\textit{x}, \textit{y}); ellipticity \textit{e} = ($\textit{a}^{2}$-$\textit{b}^{2}$)/($\textit{a}^{2}$+$\textit{b}^{2}$), where \textit{a} and \textit{b} are the semi-major and semi-minor axes, respectively; position angle \textit{$\theta$}; core radius \textit{r$_{c}$}; truncation radius \textit{r$_{cut}$}; and effective velocity dispersion \textit{$\sigma_{0}$}.  

In addition to the cluster-scale halo, the lens model assigns a mass distribution to each cluster member galaxy, to account for their contribution to the lensing potential. 
 For cluster-member selection, we obtained $grz$ photometry for all the sources in the field from the Legacy Survey Data Release 9 catalog. Cluster member galaxies were then identified using the red-sequence technique \citep{GladdersYee2000}, by comparing their $g$-$r$ color to that of the the cluster red-sequence in a color-magnitude diagram of  $g$-$r$ vs.\ $r$.  Our lens model included 29 cluster member galaxies. 
The positional parameters of the galaxy-scale halos (\textit{x}, \textit{y}, \textit{e}, \textit{$\theta$}) were fixed to the values we measured with Source Extractor \citep{BertinArnouts1996} in the LDSS3-C $r$-band imaging data. This was done in order to fully exploit the depth of the LDSS3-C data, and obtain measurements of the cluster members in the same reference frame as the lensing constraints. The galaxy-scale \textit{$\sigma_{0}$}, \textit{r$_{c}$}, and \textit{r$_{cut}$} parameters were scaled to their observed luminosity using the scaling relations in \cite{Limousin2005}.  The parameters of the cluster-scale halo, with the exception of \textit{r$_{cut}$}, were allowed to vary.  The cut radius for the cluster-scale halo was fixed to 1500~kpc because this parameter falls outside of the region where strong lensing occurs, and therefore, cannot be constrained.  We constrained the \textit{x} and \textit{y} positions of the cluster-scale halo to be within $\pm$ 10 arcseconds from the center of the brightest cluster galaxy (BCG) and set broad priors for the other free parameters.

A cluster member galaxy in the northwest (G1) contributes significant substructure to the lensing potential and likely marks an underlying dark matter subhalo.  Thus, for this galaxy, we optimized the \textit{r$_{c}$}, \textit{r$_{cut}$}, and \textit{$\sigma_{0}$}  parameters instead of adopting the values from the \cite{Limousin2005} scaling relations.  

The best-fit lens model is defined as the one that minimizes the scatter between the observed and model-predicted image locations in the image plane.  The best-fit model parameters and their 1$\sigma$ errors from the MCMC sampling of the parameter space are as follows: for the cluster-scale halo, $\Delta$R.A.\ [$''$] = -5.0$^{+3.3}_{-1.5}$; $\Delta$Decl.\ [$''$] = 
6.7$^{+1.8}_{-4.6}$; \textit{$e$}$ = $0.57$^{+0.13}_{-0.33}$;
\textit{$\theta$} = 51.9$^{+0.02}_{-2.8}$ degrees; \textit{r$_{c}$} = 80.6$^{+15.2}_{-15.9}$ kpc; \textit{r$_{cut}$} = 1500 kpc (not optimized); and  \textit{$\sigma_{0}$} = 870$^{+93}_{-112}$ km s$^{-1}$; and for G1, \textit{r$_{c}$} = 7.3$^{+1.0}_{-5.7}$ kpc; \textit{r$_{cut}$} = 29.9$^{+11.7}_{-12.7}$ kpc; and \textit{$\sigma_{0}$} = 328$^{+141}_{-25}$ km s$^{-1}$. For the scaling relations of galaxy-scale halos, the best-fit effective velocity dispersion was \textit{$\sigma_{0}^{*}$} $= 165 ^{+25}_{-23}$ km s$^{-1}$. The pivot truncation radius is poorly constrained between the priors of $20-150$ kpc  with a best fit value of \textit{r$_{cut}^{*}$} = 54.4 kpc.  Whereas the scaling relations produced \textit{$\sigma_{0}$} = 177$^{+27}_{-25}$ km s$^{-1}$ for G1, the optimized value \textit{$\sigma_{0}$} = 328$^{+141}_{-25}$ km s$^{-1}$ was significantly higher, which points to a possible existence of an underlying dark matter substructure.  A more robust investigation of this component requires more lensing constraints and is reserved for future work.  The coordinates $\Delta$R.A.\ and $\Delta$Decl.\ are listed in arcseconds measured east and north of the reference [R.A., Decl.] = [53.770260, -19.468757], the center of the BCG.

The lens model does not predict any additional lensed quasar images.  The RMS scatter in the image plane is 0$\farcs$21.  We find that the main cluster halo is generally aligned with the observed cluster light, that is, along the line connecting the few bright cluster member galaxies seen in Figure \ref{fig:imaging}.  The $x$ and $y$ positions of the main cluster halo are strongly correlated, and the lack of lensing constraints southeast of the BCG means that the position of the cluster-scale halo is under-constrained, allowing it to move freely in that direction.

Table~\ref{table:table} includes the model-predicted magnification of each of the lensed quasar images.  We used our best-fit lens model to determine the mass density projected within 250 kpc of the BCG, and measure $M(<250\,kpc) = 1.44^{+0.27}_{-0.17} \times 10^{14} M_{\odot}$.  

The statistical uncertainties on the model outputs, i.e., the mass, magnification, and time delay, were derived from a set of 100 models, sampled from the MCMC chains.  We note that the statistical modeling uncertainties underestimate the true uncertainty, as systematic uncertainties due to modeling choices, image identification, and lack of constraints can be significant (e.g., \citealt{JohnsonSharon2016}).  To obtain a better handle on systematic uncertainties and their implications, we computed several test models with different modeling assumptions, including removing constraints from the other multiply-imaged background sources, removing the faintest magnitude of cluster member galaxies, and using an NFW profile for the main halo instead of a PIEMD halo.  We also experimented with fixing the free parameters of G1 if the number of free parameters exceeded the number of constraints. We find that overall the resulting test models are consistent with our ``fiducial'' model in terms of the predicted magnification and time delay outputs, where constraints are available, albeit with larger uncertainties. As noted above, the position of the main cluster halo is poorly constrained. As a result, the best-fit cluster halo position varied significantly between test models, by more than the implied in-model statistical uncertainty. 
The results presented hereafter are reported for our ``fiducial'' lens model, while the test models were used to inform the reported uncertainties.
%\vspace{1.5cm}
\section{Predicted Time Delay}\label{sec:time_delay}
Because quasars vary in brightness over time, the time delay between the multiple lensed images of the quasar can be determined.  
Time delays between images of variable lensed sources can be used for measuring the Hubble constant \citep[e.g.,][]{Wong2020, Napier2023, Kelly2023}, and 
 are  necessary for interpreting temporal and spatial variations in the intrinsic quasar absorption lines which can then be used to probe the 3D structure of material near the quasar \citep{Misawa2016}.

Once the lensing potential is known, the time delay between images of a lensed source can be calculated from the excess arrival time surface \citep{Schneider1985}:  

\begin{equation}\label{arrival_time}
t(\vec\theta, \vec\beta) = \frac{1+z_l}{c}\frac{d_l d_s}{d_{ls}}[ \frac{1}{2} (\vec\theta-\vec\beta)^2-\psi(\vec\theta)],
\end{equation}
\vspace{-0.2in}

\noindent where $z_l$ is the lens redshift, $d_l$, $d_s$, and $d_{ls}$ are angular diameter distances from the observer to the lens, to the source, and between the lens and the source, respectively; $\vec\theta$ is the image position in the image plane; $\vec\beta$ is the unobserved source position; and $\psi(\vec\theta)$ is the gravitational lensing potential.  The last term, $\frac{1}{2} (\vec\theta-\vec\beta)^2-\psi(\vec\theta)$, is also known as the Fermat potential.  

\begin{figure}[h!]
    \centering
    \includegraphics[width=\linewidth]{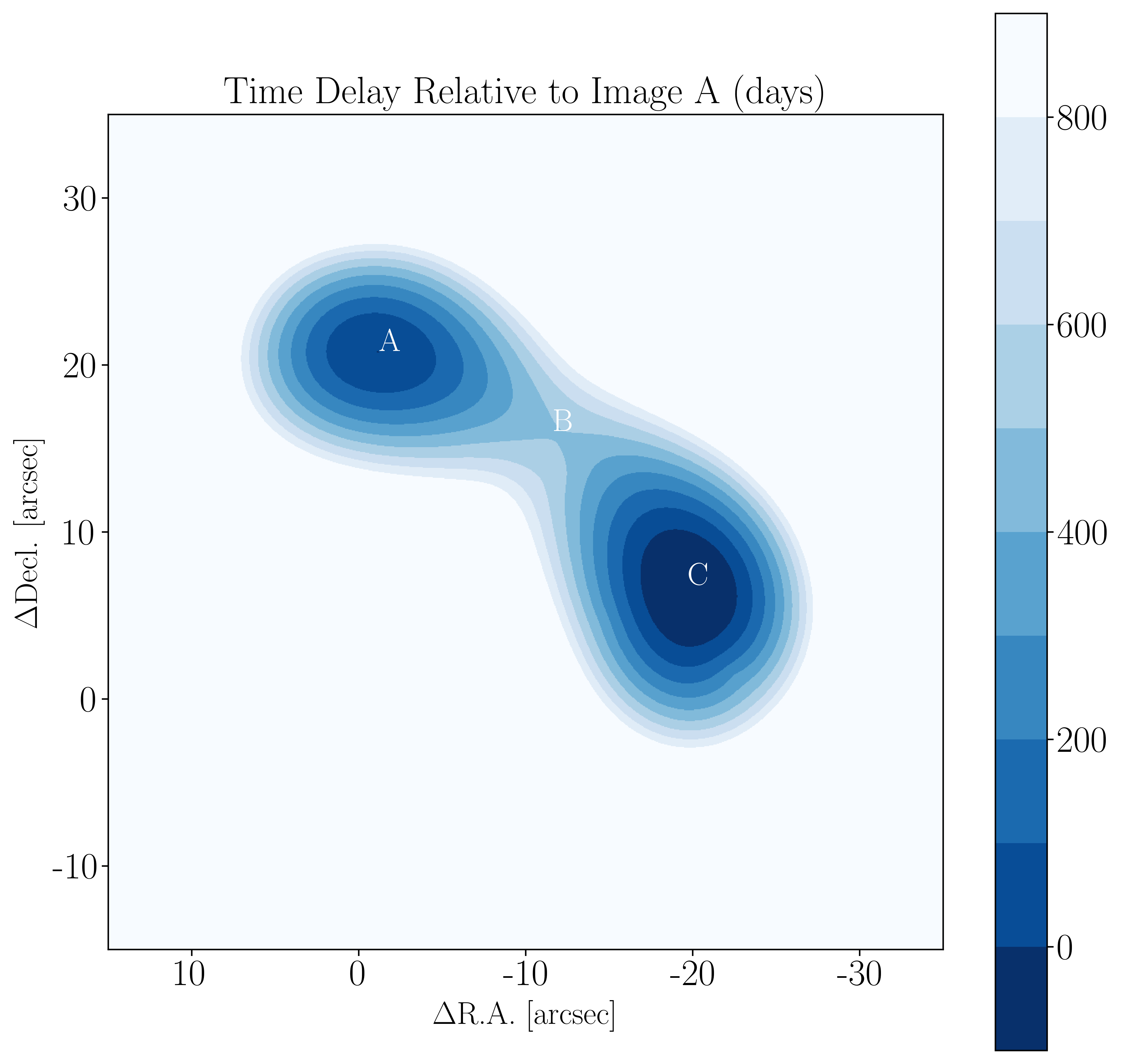}
    \caption{The excess arrival time surface for light that is emitted from the quasar source position $\beta$ at the source plane redshift \textit{z} = \qsoz\ and passes the lens plane at \textit{z} = \clusterz.  We calculated the excess arrival time, which is reported in days, relative to image A of the quasar.}
    \label{fig:excess_arrival_time}
\end{figure}

Figure~\ref{fig:excess_arrival_time} shows the predicted excess arrival time surface, in days with respect to image A of the quasar, $\Delta$t = t($\vec\theta,\vec\beta$)-t($\vec\theta_{A},\vec\beta)$, derived from the best-fit lens model.  
Multiple images of a strongly lensed background source would appear at stationary points (minimum, maximum, and saddle points) in this potential; in the case of \target, we observe two minima and a saddle point, resulting in three images of the lensed quasar.

The arrival time surface is very sensitive to small changes in the source position $\beta$.  We obtained the quasar source position $\beta$ by averaging the predicted source positions from all three lensed quasar images.  We calculated the time delays at the observed positions of the quasar images.  The best-fit model-predicted time delays are $\Delta$t$_{AB}$ = \predAB\ days and $\Delta$ t$_{AC}$ = \predAC\ days (statistical uncertainties).  To assess the systematic uncertainties, we calculated the predicted AB and AC time delays for the various test models (see Section \ref{sec:lens_model}).  The test models produced AB and AC time delays in the range of $240 < \Delta$t$_{AB} < 700$ and $-300 < \Delta$ t$_{AC} <-30$.  The statistical uncertainties likely underestimate the ``true" uncertainties by a factor of $\sim$2.  The lens model is under-constrained, and space-based imaging data would significantly improve the robustness of the time delay predictions by providing better resolution and new constraints.

\section{Initialization of Photometric Monitoring}\label{sec:photometry}
      
\begin{figure}[h!]
    \centering    
    \includegraphics[width=\linewidth]{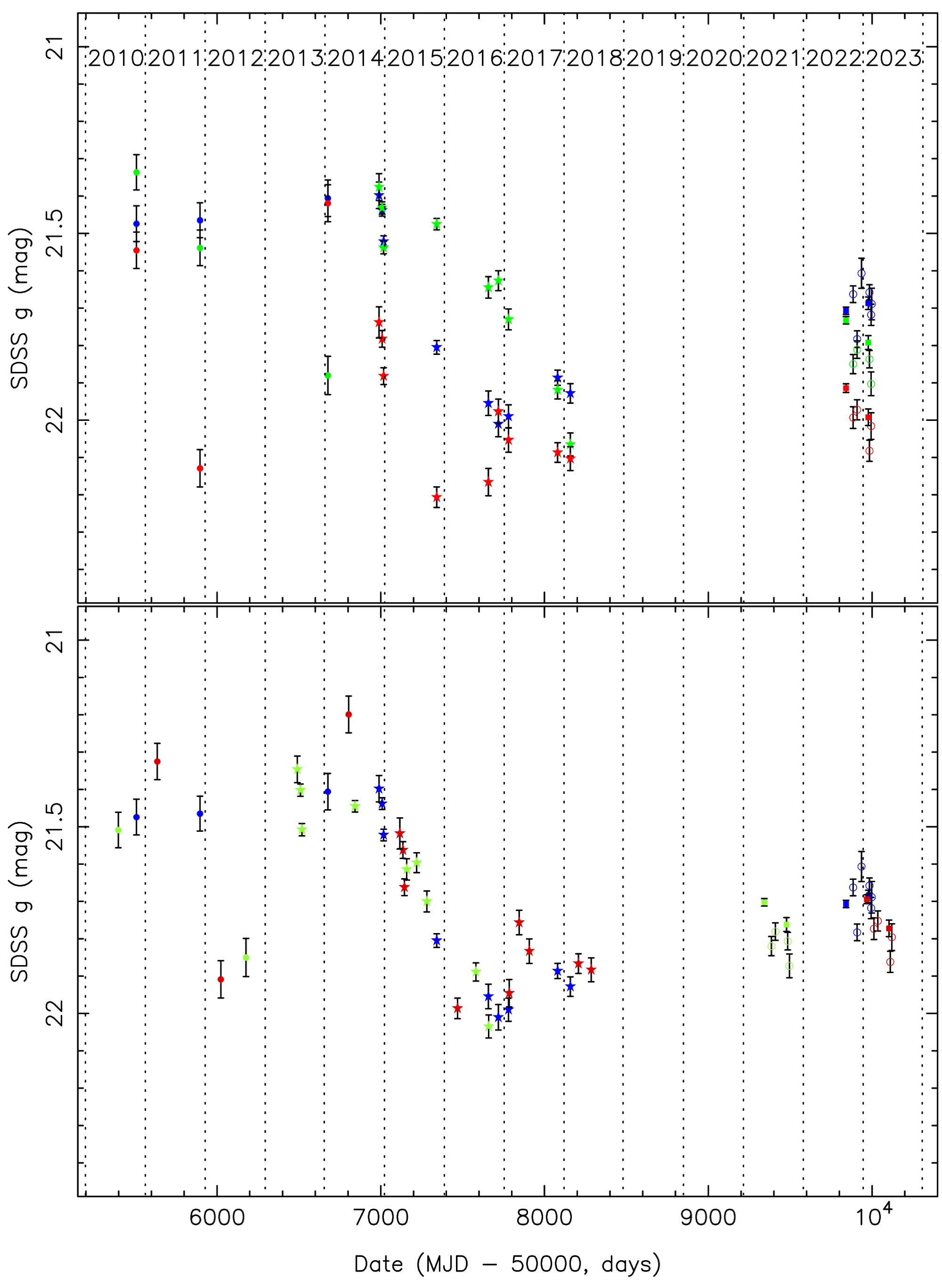}
    \caption{Top: The light curves of the quasar images A (blue), B (green), and C (red).  The data points are aperture-photometry from Pan-STARRS (filled circles), DECaLS (filled stars), the Magellan Telescopes (filled squares), and the Nordic Optical Telescope (open circles).  Bottom: The light curves for images B and C are shifted relative to image A by $\Delta$t$_{AB}=$ 499 and $\Delta$t$_{AC}=$ -127 days, the predicted time delays from the best-fit lens model.  The data points of images B and C were then shifted vertically by 0.03 mag and 0.22 mag, respectively, to align with the light curve of image A (see Section \ref{sec:photometry}).}
\label{fig:light_curves}
\end{figure}

Photometric monitoring of \target\ is essential for measuring the observed time delays between the multiple images of the lensed quasar.  
We initiated photometric monitoring of \target\ in September 2022. To date, we have obtained 8 epochs of images in the \textit{g}-band utilizing the Magellan Telescopes and the 2.56-m Nordic Optical Telescope (NOT).  
We supplemented our new observations with archival data from Pan-STARRS \citep{Flewelling2020}, spanning three epochs from late 2010 to early 2014, and from DECaLS, spanning 9 epochs between late 2014 and early 2018, to construct preliminary quasar light curves (top panel of Figure~\ref{fig:light_curves}).  The quasar images already show significant variability over time; e.g., the brightness of image A decreased by 0.6 magnitudes over a time span of two years between late 2014 and late 2016, followed by a similar decline in image B. The light curves also show that the relative order of brightness of the images has changed: The data points from 2015 and 2016 show image B as significantly brighter than image A, whereas the opposite is true in our most recent photometric data from 2023.  The DECaLS and Pan-STARRS \textit{g}-band photometry show a decline in image C around (MJD-50000) $\sim$7000d, followed by subsequent declines in image A and, finally image B, which is consistent with the model-predicted order of the time delays provided in Section~\ref{sec:time_delay}.  

We shifted the light curves for images B and C in time relative to image A by the predicted time delays from the best-fit lens model (bottom panel of Figure~\ref{fig:light_curves}).  Also, to visually match the light curve of image A, we also shifted the light curves for images B and C by 0.03 and 0.22 magnitudes, respectively, relative to image A. The predicted time delays are consistent with the preliminary light curves.  The predicted magnification ratios, however, are too uncertain to definitively say whether they are in agreement with the magnitude offsets applied to the shifted light curves.

While the data in hand, which sparsely sample a $\sim12$-year baseline, are not yet sufficient for obtaining robust time delays, the observed variability in the \textit{g}-band guarantees that such a measurement will be feasible. Higher-cadenced monitoring for this purpose will be implemented at NOT from August 2023. Furthermore, \target\ is located within the LSST footprint of the Vera C.\ Rubin Observatory which ensures high-quality light curves will be available in the near future.  

\section{Summary and Conclusions}\label{sec:conclusions}
This paper presents the discovery, spectroscopic confirmation, and preliminary photometric light curves of \target, a wide-separation lensed quasar at \z\ = \qsoz; and a preliminary strong lens model, which we computed based on the available ground-based multi-band imaging and spectroscopy. We used the best-fit lens model to make preliminary predictions of the time delays between the three images of the lensed quasar, which are qualitatively consistent with the quasar image light curves available to date. 
We demonstrated that the high variability, which is already observed in the archival and newly obtained imaging, all but guarantee that an accurate measurement of the observed time delays will be feasible once sufficient data are collected. 

Our discovery and spectroscopic confirmation of \target\ increases the small, but scientifically valuable sample of WSLQs. These unique systems enable a broad array of studies ranging from the nature of quasars and their hosts, through the properties of the foreground structures that lens them, to the expansion rate of the Universe itself. The preliminary results presented here offer a glimpse into \target, an interesting target whose analysis has just begun.

\acknowledgments
This paper is based on data gathered with the 6.5-m Magellan Telescopes located at Las Campanas Observatory, Chile. Magellan observing time for this program was granted by the time allocation committees of the University of Chicago and the University of Michigan. The authors wish to particularly thank the staff of the Las Campanas Observatory, Chile for facilitating data collection while the observatory was impacted by the global COVID-19 pandemic.

This paper is partially based on observations made with the Nordic Optical Telescope, owned in collaboration by the University of Turku and Aarhus University, and operated jointly by Aarhus University, the University of Turku and the University of Oslo, representing Denmark, Finland and Norway, the University of Iceland and Stockholm University at the Observatorio del Roque de los Muchachos, La Palma, Spain, of the Instituto de Astrofisica de Canarias. The data presented here were obtained in part with ALFOSC, which is provided by the Instituto de Astrofisica de Andalucia (IAA) under a joint agreement with the University of Copenhagen and NOT.

IRAF was distributed by the National Optical Astronomy Observatory, which was managed by the Association of Universities for Research in Astronomy (AURA) under a cooperative agreement with the National Science Foundation.

The Legacy Surveys consist of three individual and complementary projects: the Dark Energy Camera Legacy Survey (DECaLS; Proposal ID $\#$2014B-0404; PIs: David Schlegel and Arjun Dey), the Beijing-Arizona Sky Survey (BASS; NOAO Prop. ID $\#$2015A-0801; PIs: Zhou Xu and Xiaohui Fan), and the Mayall z-band Legacy Survey (MzLS; Prop. ID $\#$2016A-0453; PI: Arjun Dey). DECaLS, BASS and MzLS together include data obtained, respectively, at the Blanco telescope, Cerro Tololo Inter-American Observatory, NSF’s NOIRLab; the Bok telescope, Steward Observatory, University of Arizona; and the Mayall telescope, Kitt Peak National Observatory, NOIRLab. Pipeline processing and analyses of the data were supported by NOIRLab and the Lawrence Berkeley National Laboratory (LBNL). The Legacy Surveys project is honored to be permitted to conduct astronomical research on Iolkam Du’ag (Kitt Peak), a mountain with particular significance to the Tohono O’odham Nation.

The Pan-STARRS1 Surveys (PS1) have been made possible through contributions of the Institute for Astronomy, the University of Hawaii, the Pan-STARRS Project Office, the Max-Planck Society and its participating institutes, the Max Planck Institute for Astronomy, Heidelberg and the Max Planck Institute for Extraterrestrial Physics, Garching, The Johns Hopkins University, Durham University, the University of Edinburgh, Queen's University Belfast, the Harvard-Smithsonian Center for Astrophysics, the Las Cumbres Observatory Global Telescope Network Incorporated, the National Central University of Taiwan, the Space Telescope Science Institute, the National Aeronautics and Space Administration under Grant No. NNX08AR22G issued through the Planetary Science Division of the NASA Science Mission Directorate, the National Science Foundation under Grant No. AST-1238877, the University of Maryland, and Eotvos Lorand University (ELTE).

We thank the anonymous referee for their comments that improved the manuscript. 

\facilities{6.5-m Magellan Telescopes (Clay/LDSS3C and Baade/IMACS), 2.56-m Nordic Optical Telescope (ALFOSC)}

\software{\texttt{IRAF} \citep{Tody1986,Tody1993},  \texttt{Lenstool} \citep{Jullo2007}, \texttt{Source Extractor} \citep{BertinArnouts1996}}




%\clearpage
\bibliography{bibliography.bib}

\begin{thebibliography}{}
\expandafter\ifx\csname natexlab\endcsname\relax\def\natexlab#1{#1}\fi
\providecommand{\url}[1]{\href{#1}{#1}}
\providecommand{\dodoi}[1]{doi:~\href{http://doi.org/#1}{\nolinkurl{#1}}}
\providecommand{\doeprint}[1]{\href{http://ascl.net/#1}{\nolinkurl{http://ascl.net/#1}}}
\providecommand{\doarXiv}[1]{\href{https://arxiv.org/abs/#1}{\nolinkurl{https://arxiv.org/abs/#1}}}

\bibitem[{{Barlow} {et~al.}(1997){Barlow}, {Hamann}, \& {Sargent}}]{Barlow1997}
{Barlow}, T.~A., {Hamann}, F., \& {Sargent}, W.~L.~W. 1997, in Astronomical Society of the Pacific Conference Series, Vol. 128, Mass Ejection from Active Galactic Nuclei, ed. N.~{Arav}, I.~{Shlosman}, \& R.~J. {Weymann}, 13, \dodoi{10.48550/arXiv.astro-ph/9705048}

\bibitem[{{Bayliss} {et~al.}(2017){Bayliss}, {Sharon}, {Acharyya}, {Gladders}, {Rigby}, {Bian}, {Bordoloi}, {Runnoe}, {Dahle}, {Kewley}, {Florian}, {Johnson}, \& {Paterno-Mahler}}]{Bayliss2017}
{Bayliss}, M.~B., {Sharon}, K., {Acharyya}, A., {et~al.} 2017, \apjl, 845, L14, \dodoi{10.3847/2041-8213/aa831a}

\bibitem[{{Bertin} \& {Arnouts}(1996)}]{BertinArnouts1996}
{Bertin}, E., \& {Arnouts}, S. 1996, \aaps, 117, 393, \dodoi{10.1051/aas:1996164}

\bibitem[{{Cloonan} {et~al.}(2023){Cloonan}, {Khullar}, {Napier}, {Dahle}, {Bayliss}, {Sharon}, {Gladders}, \& {Cool-Lamps Collaboration}}]{Cloonan2023}
{Cloonan}, A., {Khullar}, G., {Napier}, K., {et~al.} 2023, in American Astronomical Society Meeting Abstracts, Vol.~55, American Astronomical Society Meeting Abstracts, 301.10

\bibitem[{{Dahle} {et~al.}(2013){Dahle}, {Gladders}, {Sharon}, {Bayliss}, {Wuyts}, {Abramson}, {Koester}, {Groeneboom}, {Brinckmann}, {Kristensen}, {Lindholmer}, {Nielsen}, {Krogager}, \& {Fynbo}}]{Dahle2013}
{Dahle}, H., {Gladders}, M.~D., {Sharon}, K., {et~al.} 2013, \apj, 773, 146, \dodoi{10.1088/0004-637X/773/2/146}

\bibitem[{{Dey} {et~al.}(2019){Dey}, {Schlegel}, {Lang}, {Blum}, {Burleigh}, {Fan}, {Findlay}, {Finkbeiner}, {Herrera}, {Juneau}, {Landriau}, {Levi}, {McGreer}, {Meisner}, {Myers}, {Moustakas}, {Nugent}, {Patej}, {Schlafly}, {Walker}, {Valdes}, {Weaver}, {Y{\`e}che}, {Zou}, {Zhou}, {Abareshi}, {Abbott}, {Abolfathi}, {Aguilera}, {Alam}, {Allen}, {Alvarez}, {Annis}, {Ansarinejad}, {Aubert}, {Beechert}, {Bell}, {BenZvi}, {Beutler}, {Bielby}, {Bolton}, {Brice{\~n}o}, {Buckley-Geer}, {Butler}, {Calamida}, {Carlberg}, {Carter}, {Casas}, {Castander}, {Choi}, {Comparat}, {Cukanovaite}, {Delubac}, {DeVries}, {Dey}, {Dhungana}, {Dickinson}, {Ding}, {Donaldson}, {Duan}, {Duckworth}, {Eftekharzadeh}, {Eisenstein}, {Etourneau}, {Fagrelius}, {Farihi}, {Fitzpatrick}, {Font-Ribera}, {Fulmer}, {G{\"a}nsicke}, {Gaztanaga}, {George}, {Gerdes}, {Gontcho}, {Gorgoni}, {Green}, {Guy}, {Harmer}, {Hernandez}, {Honscheid}, {Huang}, {James}, {Jannuzi}, {Jiang}, {Joyce}, {Karcher}, {Karkar}, {Kehoe}, {Kneib}, {Kueter-Young}, {Lan},
  {Lauer}, {Le Guillou}, {Le Van Suu}, {Lee}, {Lesser}, {Perreault Levasseur}, {Li}, {Mann}, {Marshall}, {Mart{\'\i}nez-V{\'a}zquez}, {Martini}, {du Mas des Bourboux}, {McManus}, {Meier}, {M{\'e}nard}, {Metcalfe}, {Mu{\~n}oz-Guti{\'e}rrez}, {Najita}, {Napier}, {Narayan}, {Newman}, {Nie}, {Nord}, {Norman}, {Olsen}, {Paat}, {Palanque-Delabrouille}, {Peng}, {Poppett}, {Poremba}, {Prakash}, {Rabinowitz}, {Raichoor}, {Rezaie}, {Robertson}, {Roe}, {Ross}, {Ross}, {Rudnick}, {Safonova}, {Saha}, {S{\'a}nchez}, {Savary}, {Schweiker}, {Scott}, {Seo}, {Shan}, {Silva}, {Slepian}, {Soto}, {Sprayberry}, {Staten}, {Stillman}, {Stupak}, {Summers}, {Sien Tie}, {Tirado}, {Vargas-Maga{\~n}a}, {Vivas}, {Wechsler}, {Williams}, {Yang}, {Yang}, {Yapici}, {Zaritsky}, {Zenteno}, {Zhang}, {Zhang}, {Zhou}, \& {Zhou}}]{Dey2019}
{Dey}, A., {Schlegel}, D.~J., {Lang}, D., {et~al.} 2019, \aj, 157, 168, \dodoi{10.3847/1538-3881/ab089d}

\bibitem[{{El{\'\i}asd{\'o}ttir} {et~al.}(2007){El{\'\i}asd{\'o}ttir}, {Limousin}, {Richard}, {Hjorth}, {Kneib}, {Natarajan}, {Pedersen}, {Jullo}, \& {Paraficz}}]{Eliasdottir2007}
{El{\'\i}asd{\'o}ttir}, {\'A}., {Limousin}, M., {Richard}, J., {et~al.} 2007, arXiv e-prints, arXiv:0710.5636.
\newblock \doarXiv{0710.5636}

\bibitem[{{Flewelling} {et~al.}(2020){Flewelling}, {Magnier}, {Chambers}, {Heasley}, {Holmberg}, {Huber}, {Sweeney}, {Waters}, {Calamida}, {Casertano}, {Chen}, {Farrow}, {Hasinger}, {Henderson}, {Long}, {Metcalfe}, {Narayan}, {Nieto-Santisteban}, {Norberg}, {Rest}, {Saglia}, {Szalay}, {Thakar}, {Tonry}, {Valenti}, {Werner}, {White}, {Denneau}, {Draper}, {Hodapp}, {Jedicke}, {Kaiser}, {Kudritzki}, {Price}, {Wainscoat}, {Chastel}, {McLean}, {Postman}, \& {Shiao}}]{Flewelling2020}
{Flewelling}, H.~A., {Magnier}, E.~A., {Chambers}, K.~C., {et~al.} 2020, \apjs, 251, 7, \dodoi{10.3847/1538-4365/abb82d}

\bibitem[{{For{\'e}s-Toribio} {et~al.}(2022){For{\'e}s-Toribio}, {Mu{\~n}oz}, {Kochanek}, \& {Mediavilla}}]{Fores-Toribo2022}
{For{\'e}s-Toribio}, R., {Mu{\~n}oz}, J.~A., {Kochanek}, C.~S., \& {Mediavilla}, E. 2022, arXiv e-prints, arXiv:2206.09856.
\newblock \doarXiv{2206.09856}

\bibitem[{{Gladders} \& {Yee}(2000)}]{GladdersYee2000}
{Gladders}, M.~D., \& {Yee}, H.~K.~C. 2000, \aj, 120, 2148, \dodoi{10.1086/301557}

\bibitem[{{Inada} {et~al.}(2003){Inada}, {Oguri}, {Pindor}, {Hennawi}, {Chiu}, {Zheng}, {Ichikawa}, {Gregg}, {Becker}, {Suto}, {Strauss}, {Turner}, {Keeton}, {Annis}, {Castander}, {Eisenstein}, {Frieman}, {Fukugita}, {Gunn}, {Johnston}, {Kent}, {Nichol}, {Richards}, {Rix}, {Sheldon}, {Bahcall}, {Brinkmann}, {Ivezi{\'c}}, {Lamb}, {McKay}, {Schneider}, \& {York}}]{Inada2003}
{Inada}, N., {Oguri}, M., {Pindor}, B., {et~al.} 2003, \nat, 426, 810, \dodoi{10.1038/nature02153}

\bibitem[{{Inada} {et~al.}(2006){Inada}, {Oguri}, {Morokuma}, {Doi}, {Yasuda}, {Becker}, {Richards}, {Kochanek}, {Kayo}, {Konishi}, {Utsunomiya}, {Shin}, {Strauss}, {Sheldon}, {York}, {Hennawi}, {Schneider}, {Dai}, \& {Fukugita}}]{Inada2006}
{Inada}, N., {Oguri}, M., {Morokuma}, T., {et~al.} 2006, \apjl, 653, L97, \dodoi{10.1086/510671}

\bibitem[{{Johnson} \& {Sharon}(2016)}]{JohnsonSharon2016}
{Johnson}, T.~L., \& {Sharon}, K. 2016, \apj, 832, 82, \dodoi{10.3847/0004-637X/832/1/82}

\bibitem[{{Jullo} {et~al.}(2007){Jullo}, {Kneib}, {Limousin}, {El{\'\i}asd{\'o}ttir}, {Marshall}, \& {Verdugo}}]{Jullo2007}
{Jullo}, E., {Kneib}, J.~P., {Limousin}, M., {et~al.} 2007, New Journal of Physics, 9, 447, \dodoi{10.1088/1367-2630/9/12/447}

\bibitem[{{Kelly} {et~al.}(2023){Kelly}, {Rodney}, {Treu}, {Oguri}, {Chen}, {Zitrin}, {Birrer}, {Bonvin}, {Dessart}, {Diego}, {Filippenko}, {Foley}, {Gilman}, {Hjorth}, {Jauzac}, {Mandel}, {Millon}, {Pierel}, {Sharon}, {Thorp}, {Williams}, {Broadhurst}, {Dressler}, {Graur}, {Jha}, {McCully}, {Postman}, {Borello Schmidt}, {Tucker}, \& {von der Linden}}]{Kelly2023}
{Kelly}, P.~L., {Rodney}, S., {Treu}, T., {et~al.} 2023, arXiv e-prints, arXiv:2305.06367, \dodoi{10.48550/arXiv.2305.06367}

\bibitem[{{Khullar} {et~al.}(2021){Khullar}, {Gozman}, {Lin}, {Martinez}, {Matthews Acu{\~n}a}, {Medina}, {Merz}, {Sanchez}, {Sisco}, {Kavin Stein}, {Sukay}, {Tavangar}, {Bayliss}, {Bleem}, {Brownsberger}, {Dahle}, {Florian}, {Gladders}, {Mahler}, {Rigby}, {Sharon}, \& {Stark}}]{Khullar2021}
{Khullar}, G., {Gozman}, K., {Lin}, J.~J., {et~al.} 2021, \apj, 906, 107, \dodoi{10.3847/1538-4357/abcb86}

\bibitem[{{Lemon} {et~al.}(2023){Lemon}, {Anguita}, {Auger-Williams}, {Courbin}, {Galan}, {McMahon}, {Neira}, {Oguri}, {Schechter}, {Shajib}, {Treu}, {Agnello}, \& {Spiniello}}]{lemon2023}
{Lemon}, C., {Anguita}, T., {Auger-Williams}, M.~W., {et~al.} 2023, \mnras, 520, 3305, \dodoi{10.1093/mnras/stac3721}

\bibitem[{{Lemon} {et~al.}(2019){Lemon}, {Auger}, \& {McMahon}}]{lemon2019}
{Lemon}, C.~A., {Auger}, M.~W., \& {McMahon}, R.~G. 2019, \mnras, 483, 4242, \dodoi{10.1093/mnras/sty3366}

\bibitem[{{Liesenborgs} {et~al.}(2009){Liesenborgs}, {de Rijcke}, {Dejonghe}, \& {Bekaert}}]{Liesenborgs2009}
{Liesenborgs}, J., {de Rijcke}, S., {Dejonghe}, H., \& {Bekaert}, P. 2009, \mnras, 397, 341, \dodoi{10.1111/j.1365-2966.2009.14912.x}

\bibitem[{{Limousin} {et~al.}(2005){Limousin}, {Kneib}, \& {Natarajan}}]{Limousin2005}
{Limousin}, M., {Kneib}, J.-P., \& {Natarajan}, P. 2005, \mnras, 356, 309, \dodoi{10.1111/j.1365-2966.2004.08449.x}

\bibitem[{{Martinez} {et~al.}(2023){Martinez}, {Napier}, {Cloonan}, {Sukay}, {Gozman}, {Merz}, {Khullar}, {Lin}, {Matthews Acu{\~n}a}, {Medina}, {Sanchez}, {Sisco}, {Kavin Stein}, {Tavangar}, {Gonz{\'a}lez}, {Mahler}, {Sharon}, {Dahle}, \& {Gladders}}]{Martinez2023}
{Martinez}, M.~N., {Napier}, K.~A., {Cloonan}, A.~P., {et~al.} 2023, \apj, 946, 63, \dodoi{10.3847/1538-4357/acbe39}

\bibitem[{{Misawa} {et~al.}(2013){Misawa}, {Inada}, {Ohsuga}, {Gandhi}, {Takahashi}, \& {Oguri}}]{Misawa2013}
{Misawa}, T., {Inada}, N., {Ohsuga}, K., {et~al.} 2013, \aj, 145, 48, \dodoi{10.1088/0004-6256/145/2/48}

\bibitem[{{Misawa} {et~al.}(2016){Misawa}, {Saez}, {Charlton}, {Eracleous}, {Chartas}, {Bauer}, {Inada}, \& {Uchiyama}}]{Misawa2016}
{Misawa}, T., {Saez}, C., {Charlton}, J.~C., {et~al.} 2016, \apj, 825, 25, \dodoi{10.3847/0004-637X/825/1/25}

\bibitem[{{Napier} {et~al.}(2023){Napier}, {Sharon}, {Dahle}, {Bayliss}, {Gladders}, {Mahler}, {Rigby}, \& {Florian}}]{Napier2023}
{Napier}, K., {Sharon}, K., {Dahle}, H., {et~al.} 2023, arXiv e-prints, arXiv:2301.11240, \dodoi{10.48550/arXiv.2301.11240}

\bibitem[{{Oguri}(2010)}]{Oguri2010}
{Oguri}, M. 2010, \pasj, 62, 1017, \dodoi{10.1093/pasj/62.4.1017}

\bibitem[{{Oguri} {et~al.}(2013){Oguri}, {Schrabback}, {Jullo}, {Ota}, {Kochanek}, {Dai}, {Ofek}, {Richards}, {Blandford}, {Falco}, \& {Fohlmeister}}]{Oguri2013}
{Oguri}, M., {Schrabback}, T., {Jullo}, E., {et~al.} 2013, \mnras, 429, 482, \dodoi{10.1093/mnras/sts351}

\bibitem[{{Ross} {et~al.}(2009){Ross}, {Assef}, {Kochanek}, {Falco}, \& {Poindexter}}]{Ross2009}
{Ross}, N.~R., {Assef}, R.~J., {Kochanek}, C.~S., {Falco}, E., \& {Poindexter}, S.~D. 2009, \apj, 702, 472, \dodoi{10.1088/0004-637X/702/1/472}

\bibitem[{{Schneider}(1985)}]{Schneider1985}
{Schneider}, P. 1985, \aap, 143, 413

\bibitem[{{Sharon} {et~al.}(2012){Sharon}, {Gladders}, {Rigby}, {Wuyts}, {Koester}, {Bayliss}, \& {Barrientos}}]{Sharon2012}
{Sharon}, K., {Gladders}, M.~D., {Rigby}, J.~R., {et~al.} 2012, \apj, 746, 161, \dodoi{10.1088/0004-637X/746/2/161}

\bibitem[{{Sharon} {et~al.}(2017){Sharon}, {Bayliss}, {Dahle}, {Florian}, {Gladders}, {Johnson}, {Paterno-Mahler}, {Rigby}, {Whitaker}, \& {Wuyts}}]{Sharon2017}
{Sharon}, K., {Bayliss}, M.~B., {Dahle}, H., {et~al.} 2017, \apj, 835, 5, \dodoi{10.3847/1538-4357/835/1/5}

\bibitem[{{Shu} {et~al.}(2019){Shu}, {Koposov}, {Evans}, {Belokurov}, {McMahon}, {Auger}, \& {Lemon}}]{Shu2019}
{Shu}, Y., {Koposov}, S.~E., {Evans}, N.~W., {et~al.} 2019, \mnras, 489, 4741, \dodoi{10.1093/mnras/stz2487}

\bibitem[{{Shu} {et~al.}(2018){Shu}, {Marques-Chaves}, {Evans}, \& {P{\'e}rez-Fournon}}]{Shu2018}
{Shu}, Y., {Marques-Chaves}, R., {Evans}, N.~W., \& {P{\'e}rez-Fournon}, I. 2018, \mnras, 481, L136, \dodoi{10.1093/mnrasl/sly174}

\bibitem[{{Sukay} {et~al.}(2022){Sukay}, {Khullar}, {Gladders}, {Sharon}, {Mahler}, {Napier}, {Bleem}, {Dahle}, {Florian}, {Gozman}, {Lin}, {Martinez}, {Matthews Acu{\~n}a}, {Medina}, {Merz}, {Sanchez}, {Sisco}, {Kavin Stein}, {Tavangar}, \& {Whitaker}}]{Sukay2022}
{Sukay}, E., {Khullar}, G., {Gladders}, M.~D., {et~al.} 2022, \apj, 940, 42, \dodoi{10.3847/1538-4357/ac9974}

\bibitem[{{Tody}(1986)}]{Tody1986}
{Tody}, D. 1986, in Society of Photo-Optical Instrumentation Engineers (SPIE) Conference Series, Vol. 627, Instrumentation in astronomy VI, ed. D.~L. {Crawford}, 733, \dodoi{10.1117/12.968154}

\bibitem[{{Tody}(1993)}]{Tody1993}
{Tody}, D. 1993, in Astronomical Society of the Pacific Conference Series, Vol.~52, Astronomical Data Analysis Software and Systems II, ed. R.~J. {Hanisch}, R.~J.~V. {Brissenden}, \& J.~{Barnes}, 173

\bibitem[{{Williams} {et~al.}(2021{\natexlab{a}}){Williams}, {Treu}, {Dahle}, {Valenti}, {Abramson}, {Barth}, {Dyrland}, {Gladders}, {Horne}, \& {Sharon}}]{Williams2021a}
{Williams}, P.~R., {Treu}, T., {Dahle}, H., {et~al.} 2021{\natexlab{a}}, \apj, 911, 64, \dodoi{10.3847/1538-4357/abe943}

\bibitem[{{Williams} {et~al.}(2021{\natexlab{b}}){Williams}, {Treu}, {Dahle}, {Valenti}, {Abramson}, {Barth}, {Brewer}, {Dyrland}, {Gladders}, {Horne}, \& {Sharon}}]{Williams2021b}
---. 2021{\natexlab{b}}, \apjl, 915, L9, \dodoi{10.3847/2041-8213/ac081b}

\bibitem[{{Wong} {et~al.}(2020){Wong}, {Suyu}, {Chen}, {Rusu}, {Millon}, {Sluse}, {Bonvin}, {Fassnacht}, {Taubenberger}, {Auger}, {Birrer}, {Chan}, {Courbin}, {Hilbert}, {Tihhonova}, {Treu}, {Agnello}, {Ding}, {Jee}, {Komatsu}, {Shajib}, {Sonnenfeld}, {Blandford}, {Koopmans}, {Marshall}, \& {Meylan}}]{Wong2020}
{Wong}, K.~C., {Suyu}, S.~H., {Chen}, G. C.~F., {et~al.} 2020, \mnras, 498, 1420, \dodoi{10.1093/mnras/stz3094}

\bibitem[{{Zhang} {et~al.}(2022){Zhang}, {Manwadkar}, {Gladders}, {Khullar}, {Dahle}, {Napier}, {Mahler}, {Sharon}, {Matthews Acu{\~n}a}, {Ashmead}, {Cerny}, {Remolina Gonz{\`a}lez}, {Gozman}, {Levine}, {Marohnic}, {Martinez}, {Merz}, {Pan}, {Sanchez}, {Sierra}, {Sisco}, {Sukay}, {Tavangar}, \& {Zaborowski}}]{Zhang2022}
{Zhang}, Y., {Manwadkar}, V., {Gladders}, M.~D., {et~al.} 2022, arXiv e-prints, arXiv:2212.06902, \dodoi{10.48550/arXiv.2212.06902}

\end{thebibliography}
\bibliographystyle{aasjournal}

\begin{table}[h!]
\begin{center}
\begin{tabular}{c c c c c c c}
 \hline
 ID & R.A. [J2000] & Decl. [J2000] & \textit{z} & $\mu$ &$\Delta$t (days) & rmsi [\arcsec] \\
 \hline
  QSO Image A & 53.769891 & -19.462959 & 3.27 & 5.3$^{+3.4}_{-1.8}$ & \nodata & 0.09 \\
  QSO Image B & 53.766807 & -19.464290 & 3.27 & 4.5$^{+3.2}_{-1.7}$ & \predAB & 0.28  \\
  QSO Image C & 53.764431 & -19.466858 & 3.27 & 5.3$^{+4.5}_{-1.5}$ & \predAC & 0.16  \\
  2.1 & 53.765563 & -19.463420 & 2.91$^{+0.56}_{-0.28}$ & & & 0.19 \\
  2.2 & 53.764377 & -19.464362 & 2.91$^{+0.56}_{-0.28}$ & & & 0.15 \\
  2.3 & 53.768347 & -19.461810 & 2.91$^{+0.56}_{-0.28}$ & & & 0.13 \\
  3.1 & 53.765909 & -19.466316 & 1.29$^{+0.22}_{-0.14}$ & & & 0.04 \\
  3.2 & 53.767522 & -19.464820 & 1.29$^{+0.22}_{-0.14}$ & & & 0.35 \\
  3.3 & 53.768146 & -19.464550 & 1.29$^{+0.22}_{-0.14}$ & & & 0.28 \\
  4.1 & 53.769297 & -19.464671 & 1.25$^{+0.21}_{-0.12}$ & & & 0.20 \\
  4.2 & 53.766640 & -19.466771 & 1.25$^{+0.21}_{-0.12}$ & & & 0.24 \\
\end{tabular}
\caption{The positions and redshifts of the multiple-image systems in the \target\ field.  The quasar images have a spectroscopic redshift \textit{z}. Redshifts with error bars indicate the median lens model-optimized redshift and the 1$\sigma$ confidence interval.  For the quasar images, the lens model-predicted magnifications and time delays and their 1$\sigma$ errors from the MCMC sampling of the parameter space are listed.  The RMS scatters in the image plane between the observed image positions and the model-predicted positions are listed in the last column.  The ground-based lens model is uncertain and the statistical error bars do not represent the true uncertainties, which are likely underestimated (e.g., see Figure 8 in \cite{Sharon2012})}.
\label{table:table}
\end{center}
\end{table}

\end{document}